# Comparison and Analysis of Twist Pitch Length Test Methods for ITER Nb$_3$Sn and NbTi Strands[*]

LIU Fang[1)]  LONG Feng  CHEN Chao  LIU Bo  WU Yu  and  LIU Huajun

Institute of Plasma Physics, Chinese Academy of Science (ASIPP), Hefei, 230031, People's Republic of China.

**Abstract**

A twisted multifilamentary structure is needed for Nb$_3$Sn and NbTi strands to be used in the International Thermonuclear Experimental Reactor (ITER) magnets. The Cable-In-Conduit Conductor (CICC) for the coils of ITER relies on twisted, multifilament, Nb$_3$Sn and NbTi based composite strands. As important parameters for the superconducting strand's design and production, the twist pitch length and direction of strands must meet the requirements according to ITER Procurement Arrangement (PA) and this must be verified. The technical requirements are 15mm±2mm for twist pitch length and right hand twist for direction. The strand twist pitch and the twist direction can be measured on straight sections of strand, which is recognized by the repetition of filament bundles or by the angle of the filaments. Several test methods and results are described and compared in this paper. The accuracy, uncertainty and feasibility of different methods are analyzed and recommended measurement methods are proposed for ITER strands verification.

**PACS: 0007**

## 1. Introduction

ITER is designed to be the world's largest experimental fusion facility. The CICC for the Coils of ITER relies on twisted, multifilament, Nb$_3$Sn and NbTi based composite strands [1,2]. Two types of NbTi based composite strands (type 1 and type 2) are used for three different cable layouts/conductor types. Two strand manufacturing routes for Nb$_3$Sn-based composite strands are typically used for ITER: the Bronze route (BR) and the Internal Tin route (IT). Both routes rely on precursors of Nb$_3$Sn, which are assembled and transformed into strands.

The twist pitch mainly influences the coupling loss of the conductor, which is generated within the matrix metal by inter-filament currents that are induced under a time-varying magnetic field excitation [3]. The technical requirements are 15mm±2mm for twist pitch length $T_P$ and right hand twist for direction for ITER Nb$_3$Sn and NbTi strand. As important parameters for the superconducting strands design and production, the twist pitch $T_P$ and twist direction of strands must meet the requirements according to ITER PA and this must be verified.

Several test methods can be used for $T_P$ determination. In this paper, we present three methods (repetition of filament bundles, angle of the filaments and rotation) for $T_P$ measurement and compare them. Two of them (repetition of filament bundles, angle of the filaments) which are similar to the methods instructed in ITER PA are described in detail. A new method, i.e. rotation method, which is suitable for NbTi strand is also described. Considering the characteristics of NbTi and Nb$_3$Sn strands, different methods will be recommended.

## 2. Strands

For NbTi strand, the multifilament composite strand is made up of NbTi filaments embedded in a high purity Cu matrix. The last stage of strand fabrication starts from the assembly of a final composite billet consisting of NbTi rods wrapped with Nb foils in a Cu matrix. The billet is drawn down in multiple passes to the final strand diameter. The drawing passes are interleaved with high temperature heat treatments to favor the production of α-Ti precipitates and achieve high pinning forces. A micrograph of the transverse cross-section of the NbTi type 2 strand, which was fabricated

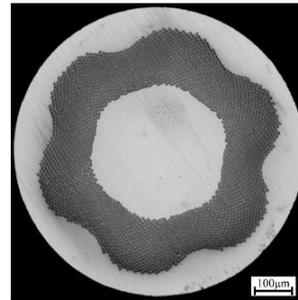

by the Western Superconducting Technology Company (WST), is shown in Fig. 1.

Fig.1. Transverse cross-section micrograph of the WST NbTi (type 2) strand used for ITER conductor. (Images by courtesy of Gao Huixian)

For Nb$_3$Sn strand, the un-reacted multifilament composite strand is made up of Nb filaments embedded in a high tin bronze matrix (BR route) or in a mixed copper and tin matrix (IT route), surrounded by a barrier which separates the multifilament zones from a stabilizing copper sheath. Ti or Ta sources may be added to or alloyed with the Nb to create a ternary alloy [4]. The last stage of strand fabrication starts from the assembly of a final composite billet, consisting of Nb rods in a bronze or a Cu matrix, Nb or Ta barriers, stabilizing Cu in the form of an outer can and Sn elements in the case of IT strands. The billet is drawn down in multiple passes to the final strand diameter. In the case of the BR route, the drawing passes are interleaved with annealing heat treatments at moderate temperatures to maintain the workability of the bronze. The micrographs of the transverse cross-section of the

[*] Supported by the Equipment Function Developing and Technical Innovation project of the Chinese Academy of Sciences under Grant Y25ZB13291.
  1)  fangliu@ipp.ac.cn

Nb₃Sn strands, which were fabricated by WST, are shown in Fig. 2.

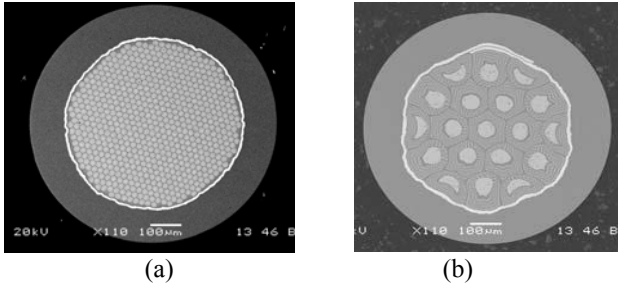

(a) (b)

Fig. 2. Transverse cross-section of Nb₃Sn strand. (a) the WST BR Nb₃Sn strand; (b) the WST IT Nb₃Sn strand used for ITER conductor. (Images by courtesy of Gao Huixian)

The technical requirements for ITER strands are shown in Table 1.

Table 1. Technical requirements for ITER strands

| Item | Requirement | | |
|---|---|---|---|
| Superconductor type | Nb₃Sn | NbTi type1 | NbTi type2 |
| Strand diameter | 0.820 ± 0.005 mm | 0.730 ± 0.003 mm | 0.730 ± 0.003 mm |
| Twist pitch | 15 ± 2 mm | 15 ± 2 mm | 15 ± 2 mm |
| Twist direction | right hand twist | right hand twist | right hand twist |
| plating thickness | 2.0 +0/-1 μm | 2.0+0/-1 μm | 2.0+0/-1 μm |
| Cu-to-non-Cu volume ratio | 1.0 ± 0.1 | 1.6 ± 0.05 | 2.35± 0.1 |
| Filament diameter, $D_{fil}$ | -- | ≤ 8 μm | ≤ 8 μm |
| Inter-filament spacing, $S_{fil}$ | -- | ≥ 1 μm | ≥ 1 μm |

## 3. Test methods

There different methods for $T_P$ measurement were introduced and analyzed.

### 3.1 Method A: repetition of filament bundles

The twist pitch can be measured on polished longitudinal sections of strand. A short (~25 mm long) section of strand is flat embedded into resin and it is carefully polished to the middle, i.e. the axis, of the strand. The pitch can be estimated looking at the outer layer of filaments emerging from the matrix in the longitudinal section.

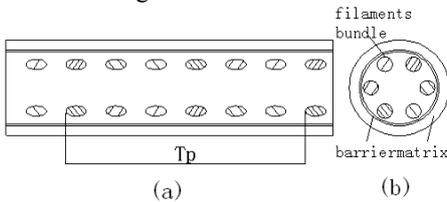

Fig. 3. Schematic diagram of $T_P$ measurement by repetition of filament bundles

The technique involves counting the number of sub-bundles around the outer layer of the filament pack (let's say there are n bundles), and then selecting every n$^{th}$ bundle in the longitudinal cross-section. The schematic diagram is shown in Fig. 3. For a wire with 6 filament bundles (see Fig. 3 (b)), the longitudinal cross-section is shown in Fig. 3(a) with twist direction right hand. The twist pitch $T_P$ is recognized by the repetition of filament bundles [5].

### 3.2 Method B: angle of the filaments

By method B, the twist pitch can be measured from the twist angle of filaments. The sample is prepared by chemically etching away the stabilizing Cu with the two ends of samples fixed before etching to make the sample be straight through the measuring process. The photos can be taken by SEM.

The schematic diagram is shown in Fig. 4. The angle $\alpha$ is the twist angle of the filament, i.e. the tangent angle. The twist pitch $T_P$ is calculated by Eq. (1).

$$T_P = \pi D \times \cot \alpha \qquad (1)$$

where $D$ is the filamentary diameter.

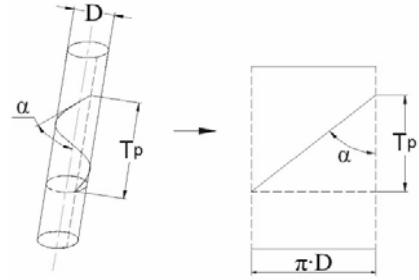

Fig. 4. Schematic diagram of $T_P$ measurement by angle of the filaments

The twist angle can be obtained by image analysis of SEM. Numerous (tens of) angles are obtained and measured. These results are averaged to minimize random error in the final result. The region for angle determination has large effects on measurement results. The angle measured from the photo is not exactly the tangent angle of the filament. Twist angle measurement should be made on the middle (or center) section of the sample for best accuracy. The relationship between the twist angle $\alpha$ and measured twist angle $\beta$ will be analyzed below.

The accurate twist angle $\alpha$ is the tangent angle of the helix (see Fig.5). The helix equation is given in Eq. (2).

$$\begin{cases} x = d \cos \theta \\ y = d \sin \theta \\ z = b\theta \end{cases} \qquad (2)$$

Where, $d$ is radius of the cross section, i.e. the filamentary radius, $\theta$ is the rotation angle, and $2\pi b = T_p$. So $b$ can be presented by Eq. (3).

$$b = \pi D \times \cot \alpha / 2\pi = d \cot \alpha \qquad (3)$$

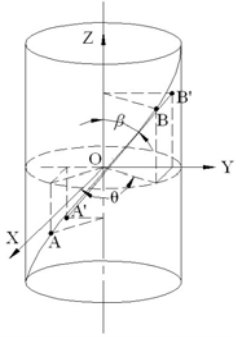

Fig.5. Helix

As shown in Fig.5, two points A and B are located on the helix. The projection of A and B on plane YOZ are A' and B', respectively. When the region from point A to B is used for twist angle determination, the measured twist angle $\beta$ is the angle between Z-axis and line A'B'. And the measured twist pitch $T_{Pm}$ is presented by Eq. (4).

$$T_{Pm} = \pi D \times \cot \beta \tag{4}$$

As shown in Fig.5, projection points A' and B' can be presented by Eq. (5), respectively,

$$\begin{cases} \vec{r_{A'}} = d \sin \theta_A \vec{j} + b\theta_A \vec{k} \\ \vec{r_{B'}} = d \sin \theta_B \vec{j} + b\theta_B \vec{k} \end{cases} \tag{5}$$

Then the measured twist angle $\beta$ can be presented by Eq. (6).

$$\cot \beta = \frac{b\theta_B - b\theta_A}{d \sin \theta_B - d \sin \theta_A} \tag{6}$$

Put Eq. (4) into Eq. (6), then the ratio of measured twist pitch $T_{Pm}$ to $T_P$ can be presented by Eq. (7)

$$\frac{T_{Pm}}{T_P} = \frac{\theta_B - \theta_A}{\sin \theta_B - \sin \theta_A} \tag{7}$$

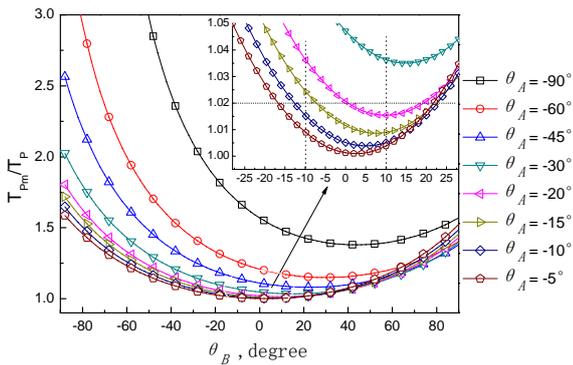

Fig. 6. Relationship between the ratio of $T_{Pm}$ to $T_P$ and the rotation angle

The rotation angles $\theta$ of the points are from -90 degree to 90 degree. When we set the rotation angle of point A $\theta_A$ as a constant, the relationship between the ratio of $T_{Pm}$ to $T_P$ and the rotation angle of point B $\theta_B$ is shown in Fig. 6.

As shown in Fig. 6, the deviation of $T_P$ caused by the angle determination can be less than 2% when we set the rotation angle $\theta$ from -10 degree to 10 degree (see Fig.10) as the angle measured region.

### 3.3 Method C: Rotation

The strand twist pitch and the twist direction can be measured on straight sections of strand, stretched on a fixture holding the ends. The Copper stabilizer of an about 200 mm long section of the strand was etched by $HNO_3$ solution to remove the copper and bronze.

The schematic diagram is shown in Fig. 7. The test procedure was as follows :1), Fix the sample with a holder; 2), etch the middle part of the sample and measure the etching length L; 3), Rotate one end of the sample clockwise till the filaments become untwisted.

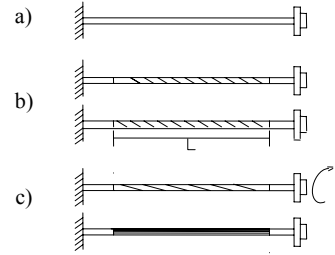

Fig.7. The $T_P$ test process by rotation method

The untwisting of strand sample will worsen the surface condition since more filaments will pop up and be separated from the bundle. It is impossible to just distinguish if the filaments are twisted or untwisted by picture or eyesight. A thin smooth sheet may be used to select some filaments and divide from others carefully. Make the sheet go from one end to another along the etched section of the sample. Adjust the rotation angle to make the sheet be at same position of the two ends and make the selected filaments have no twist from others, then record the rotating angle $\delta$ or number of turns $n$.

The twist pitch $T_P$ can be calculated by Eq. (8).

$$T_P = L / n = L / (\delta / 360) \tag{8}$$

## 4. Test Results and discussion

### 4.1 Nb$_3$Sn strand

Methods B and C are useful for both IT and BR Nb$_3$Sn strands. Method A is only valid and reliable for strands consisting of highly apparent repeating pattern in outermost layer of filament bundle. So for the Nb$_3$Sn strands in Fig 2, all the three methods could be used to test the $T_P$.

By method A, the longitudinal cross-section of IT Nb$_3$Sn for the 2$^{nd}$ benchmarking strand, whose cross section is shown in reference [6], is shown in Fig. 8. The outside filament bundles are total of 12. The Tin core is quite easily recognized. The length along the 12 outside Tin cores is $T_P$.

The longitudinal cross-section of BR Nb$_3$Sn for the 1$^{st}$ benchmarking strand, whose cross section was shown in reference [7], is shown in Fig. 9. The outside filament bundles have 4-1-4-1…construction (see Fig.9). The twist pitch is recognized by the 4-1-4-1…repetition of filament bundles [7]. For WST BR $Nb_3Sn$ strand, the outside filament bundles have 8-1-1-8-1-1-8… construction. The results of the two rounds of world-wide benchmarking are shown in Table 2.

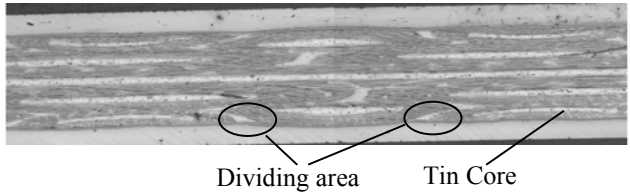

Fig. 8. Micrograph of a longitudinal cross-section for 2nd benchmarking IT $Nb_3Sn$

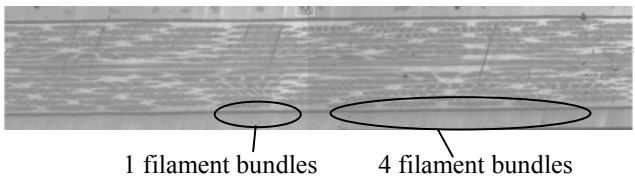

Fig. 9. Micrograph of a longitudinal cross-section for 1st benchmarking BR $Nb_3Sn$.

By method B, after removing the copper matrix, the tangent angle can be measured by image analysis of a digital photomicrograph. The angle is between the filament and strand axial direction, see Fig. 10. The strand axial direction can be determined by an un-etched strand which is parallel to the measuring strand with special holder. The filamentary diameter $D$ is determined by the cross section graph. $D$ of each sample should be measured for more accurate testing. This value will have a small deviation less than 1%. Numerous angles will be averaged to eliminate the local variability.

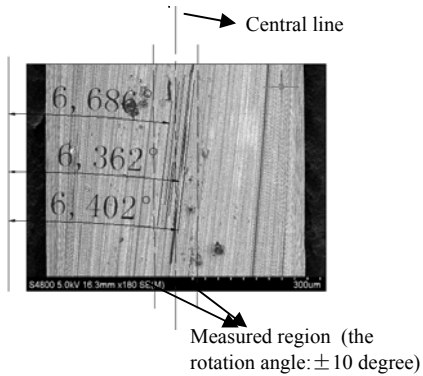

Fig. 10. Microphotograph of tangent angles of $Nb_3Sn$ strand.

For test method C, the sample holder should be designed to fix the sample with one end free of rotation. The barrier won't be removed by $HNO_3$ solution. So unlike from NbTi, the etching processes require two stages.

After etching the copper matrix, the barrier should be removed. For WST ITER IT strand, Huixian Gao from WST suggested a reliable procedure to remove the Ta barrier just by metallographic sandpaper without significant brokenness of the filaments. Chemical etching with hydrofluoric solution may be another option. But from now on, we failed to remove the Ta barrier using hydrofluoric solution.

Then the bronze or copper in the filament region can be removed by $HNO_3$ solution. From Table 2, the results by rotation method of $Nb_3Sn$ sample for the 1st round of world wide benchmarking fit quite well with the average value.

**4.2 NbTi strand**

For NbTi strand, repetition of filament bundles method is not available because the filaments are hard to be divided.

The other two methods have been compared.

For test method B, after removing the copper matrix, the NbTi filaments region will have comparatively large distortion. This may be caused by the movement of the filaments after copper removal. Without Cu matrix, some outer filaments will loose from its bundle and the twist angle will be change. To minimize this effect, enough number of measurements should be made.

Another error source is from the filament diameter determination. The deviation from the largest to smallest part is around 5% for the sample in Fig 1. Unlike from $Nb_3Sn$, there is no barrier existing, and the inner filaments may be chosen for more accurate angle determination during the angle determination.

NbTi strand is comparatively adaptable to use method C: rotation method. Both the accuracy and the efficiency are much higher than using other ways. Sample holder should be designed to fix the sample with one end free of rotation.

From Table 2, the results by rotation method of NbTi sample for the 2nd world wide benchmarking fit quite well with the average value.

Table 2 $T_P$ Test results with different methods

| Sample | | Twist Pitch, mm | | | Mean Value [a] |
|---|---|---|---|---|---|
| | | A | B | C | |
| Benchmarking | BR-$Nb_3Sn$1 | 15.8 | 16.1 | 15.8 | 15.9 mm |
| | IT-$Nb_3Sn$1 | 15.5 | 15.6 | 15.2 | / |
| | NbTi | / | 13.5 | 14.7 | 15.1 mm |
| Verification [b] | IT-$Nb_3Sn$1 | 15.9 | 15.7 | 16.0 | / |
| | NbTi1 | / | 14.1 | 13.4 | / |
| | NbTi2 | / | 14.5 | 13.7 | / |

[a] The mean value is from the results of the two rounds world-wide benchmarking for ITER strands [6, 7].
[b] Verification samples to be measured by Chinese Domestic Agency (CNDA)

**5. Conclusion**

Twist pitch plays a key role for the ITER superconducting strand's performance. Several methods can be used for twist pitch determination. Considering the characteristics of ITER $Nb_3Sn$ and NbTi strands, three methods were introduced and analyzed in this paper.

For $Nb_3Sn$ strands, all of the three methods were used to measure the twist pitch. Method B is recommended because it is a good compromise between cost and accuracy for strands verification for ITER conductor fabrication. For method A, the sample preparation may cost too much time although it has very high accuracy. In order for method C to be suitable, a reliable Ta barrier removal method must be implemented.

For NbTi strands, two methods may be useful including the filament angle method and the rotation method. Using method B, the systematic error can be much larger for NbTi strand than for $Nb_3Sn$. Additionally, it is hard to control the error because there is no barrier for the NbTi strands. Considering both cost and accuracy for strands verification, the rotation method is recommended for NbTi strands, and the error can be controlled to be a low level.

**Acknowledgment**

We thank Sander Wessel from Twente University and Huixian Gao from Western Superconducting Technologies company (WST) for many useful discussions.

**References**

[1] Mitchell N., Bessette D., Gallix R., Jong C., Knaster J., Libeyre P., Sborchia C. and Simon F., "The ITER magnet system," *IEEE Trans. Appl. Supercond.*, 2008, 18:435.
[2] N. Mitchell, P. Bauera and D. Bessette, "Status of the ITER magnets", *Fusion Eng. Des.*, 2009, 84:113.
[3] Yukikazu Iwasa, *Case Studies in Superconducting Magnets*, Plenum Press, New York, 1994.261-263.
[4] A. Godeke, *Performance Boundaries in $Nb_3Sn$ Superconductors*, Ph.D. dissertation [Dissertation], Twente University, The Nertherlands, 2005:48.
[5] http://www.utwente.nl/tnw/ems/Research/
[6] I. Pong, M. C. Jewell, B. Bordini et al., "World-wide benchmarking of Internal Tin $Nb_3Sn$ strand and NbTi strand Test facilities", *IEEE Trans. Appl. Supercond.*,2012,22(3):4802606.
[7] M. C. Jewell, T. Boutboul, L. R. Oberli, F. Liu, Y. Wu, A. Vostner et al., "World-wide benchmarking of ITER $Nb_3Sn$ strand test facilities", *IEEE Trans. Appl. Supercond.*, 2010, 20( 3):1500-1503.